# Anomalous resistivity and the electron-polaron effect in the two-band Hubbard model with one narrow band

M. Yu. Kagan[1], V.V. Val'kov[2]

[1]P.L. Kapitza Institute for Physical Problems, Kosygina st. 2, 119334 Moscow, Russia,
kagan@kapitza.ras.ru

[2]Kirenskii Institute of Physics, Akademgorodok 50, building 38, 660036 Krasnoyarsk, Russia,
vvv@iph.krasn.ru

**Abstract.** We search for anomalous normal and superconductive behavior in the two-band Hubbard model with one narrow band. We analyze the influence of electron-polaron effect and Altshuler-Aronov effect on effective mass enhancement and scattering times of heavy and light components in the clean case. We find anomalous behavior of resistivity at high temperatures $T > W_h^*$ both in 3D and 2D situation. The SC instability in the model is governed by enhanced Kohn-Luttinger effect for p-wave pairing of heavy electrons via polarization of light electrons.

## 1. Introduction.

The two-band model plays an important role both in the physics of conventional multiband s-wave superconductors like Nb [1,2] and in the physics of high-$T_c$ materials [3]. It can be useful also for the description of anomalous normal and superconducting properties in different unconventional superconductors such as ruthenates $Sr_2RuO_4$, $MgB_2$, new superconductors (SC) based on FeAs layers such as $BaFe_2(As_{1-x}P_x)_2$ [4,5], layered semimetals, dichalcogenides and superlattices, organic superconductors et al. [6-8]. At the recent conference on stripes and high Tc superconductivity in Rome the fermion-boson multiband SC and Bose-BCS crossover was also analyzed in [9,10]. In this context we could mention as well [11,12]. In the present paper we consider two-band fermionic Hubbard model with one narrow band [13,14]. This model is a very rich one. It describes adequately mixed valence systems such as uranium-based HF and possibly also some other novel superconductors and transition-metal systems with orbital degeneracy such as complex magnetic oxides in optimally doped case. Moreover it contains such highly nontrivial effect as Electron Polaron Effect [15,16] in the homogeneous state. Let us verify this model with respect to marginality [17-19] and anomalous resistivity characteristics.

## 2. Two-band Hubbard model.

In real space the Hamiltonian of the two-band Hubbard model reads:

$$\hat{H}' = -t_h \sum_{<ij>\sigma} a^+_{i\sigma} a_{j\sigma} - t_L \sum_{<ij>\sigma} b^+_{i\sigma} b_{j\sigma} - \varepsilon_0 \sum_{i\sigma} n^\sigma_{ih} - \mu \sum_{i\sigma}(n^L_{i\sigma} + n^h_{i\sigma}) + U_{hh} \sum_i n^\uparrow_{ih} n^\downarrow_{ih} + U_{LL} \sum_i n^\uparrow_{iL} n^\downarrow_{iL} + \frac{U_{hL}}{2} \sum_i n_{iL} n_{ih} \quad (1)$$

We consider low density $(n_h + n_L)d^D \ll 1$ and strong-coupling limit $U_{hh} \sim U_{hL} \sim U_{LL} \gg W_L \gg W_h$ of this model, $\mu$ is chemical potential. The band structure is presented in Fig. 1.

## 2.1 Electron-polaron effect.

We call the electron-polaron effect (EPE) the non-adiabatical part of many-particle wave function which describes the heavy particle dressed in a cloud of virtual electron-hole pairs of light particles. Nonadiabaticity of the cloud in some energy interval manifests itself when the heavy particle moves from one elementary cell to a neighboring one. Formally EPE is connected with interband Hubbard interaction $U_{hL}$. In the second order of perturbation theory

$$m_h^* / m_h = Z_h^{-1} = 1 + b \ln \frac{m_h}{m_L}, \qquad (2)$$

where $b = 2f_0^2$, $Z_h^{-1} = 1 - \left.\frac{\partial \Sigma_{hL}(\omega, \varepsilon_q)}{\partial \omega}\right|_{\omega \to 0}$; (3)

$Z_h$ is Z-factor of heavy particle. In Born approximation $f_0 = U_{hL} \nu_L(\varepsilon_F)$. In more general case of low density and strong Hubbard interaction

$$U_{hL} > W_L : f_0 = \frac{2dp_F}{\pi} - \qquad (4)$$

Galitskii gas parameter in 3D, $d$ is intersite distance. In 2D

$$f_0 = \frac{1}{2\ln 1/dp_F} \qquad (5)$$

- gas-parameter of Bloom. Generally (after collecting of the polaron exponent) the effective mass reads

$$\frac{m_h^*}{m_h} \sim \left(\frac{m_h}{m_L}\right)^{b/(1-b)}$$ [15,16]. In the unitary limit the polaron exponent $b$ can reach the value of ½ and

thus $\frac{m_h^*}{m_h} \sim \frac{m_h}{m_L}$. Accordingly $\frac{m_h^*}{m_L} \sim \left(\frac{m_h}{m_L}\right)^2$ and if we start with $\frac{m_h}{m_L} \sim 10$ in LD approximation for example we can finish with $\frac{m_h^*}{m_L} \sim 100$ due to many-body effects. Thus EPE can possibly explain the origin of a heavy mass in uranium-based HF.

## 2.2 Tendency towards phase separation.

Let us consider other mechanisms of mass-enhancement. The EPE is connected with Z-factor of heavy particle

(with $\left.\frac{\partial \Sigma_{hL}(\omega, \varepsilon_{\bar{q}})}{\partial \omega}\right|_{\omega \to 0}$ ).

However in 3D-case momentum dependence of heavy-light self energy

$$\left.\frac{\partial \Sigma_{hL}(\omega, \varepsilon_{\vec{q}})}{\partial \varepsilon_{\vec{q}}}\right|_{q \to p_F} \qquad (6)$$

also becomes very important. Hence as it is shown in [13,14] the full expression for $m^*_h/m_h$ in the second order of perturbation theory reads

$$\frac{m^*_h}{m_h} = 1 + b \ln \frac{m_h}{m_L} + \frac{b}{18} \frac{m_h n_h}{m_L n_L} \qquad (7)$$

and possess the additional term which is linear in the bare mass-ration $m_h/m_L$. If in LD approximation $m_h \sim 10 m_L$, then this term becomes dominant over EPE contribution $\sim \ln \frac{m_h}{m_L}$ for large density mismatch $n_h \geq 5 n_L$. It is very interesting that in 3D the same parameter $b \frac{m_h n_h}{m_L n_L} \geq 1$ governs the tendency towards phase-separation in the two-band model yielding negative partial compressibility

$$\chi_{hh}^{-1} \sim c_h^2 \sim (n_h / m_h)(\partial \mu_h / \partial n_h) \qquad (8)$$

where $\mu_h$ is chemical potential of the heavy particle. This result is in qualitative agreement with predictions of mean-field type variational analysis [20]. Note that in 2D case the contribution to $m^*_h$ from $\frac{\partial \Sigma_{hL}}{\partial \varepsilon_{\vec{q}}}$ and the tendency towards phase-separation are absent due to specific form of polarization operator [13,14].

## 3. Transport properties

### 3.1 Resistivity in the homogeneous case in 3D.

Exact solution of coupled kinetic equations with an account of umklapp processes yields for $p_{Fh} \sim p_{FL} \sim p_F \sim 1/d$ and low temperature $T < W^*_h < W_L$ for the inverse scattering times: $1/\tau_L \sim 1/\tau_{Lh} \sim f_0^2 \frac{T^2}{W^*_h} \frac{m_h}{m_L}$; $1/\tau_h \sim 1/\tau_{hL} \sim f_0^2 \frac{T^2}{W^*_h}$. This behavior corresponds to Landau Fermi-liquid picture. Accordingly for the conductivities we have: $\sigma_h \sim \sigma_{hL} \sim \sigma_L \sim \sigma_{Lh} \sim \frac{\sigma_{min}}{b} \left(\frac{W^*_h}{T}\right)^2$ at low temperatures $T < W^*_h$. Thus the resistivity

$$R \sim \frac{1}{(\sigma_h + \sigma_L)} \sim \frac{b}{\sigma_{min}} \left(\frac{T}{W^*_h}\right)^2, \qquad (9)$$

where $\sigma_{min} = \frac{e^2 p_F}{\hbar}$ is minimal Mott-Regel conductivity in 3D. At high temperatures $T > W^*_h$ the inverse scattering times read: $1/\tau_L \sim 1/\tau_{Lh} \sim bW_L$, $1/\tau_h \sim 1/\tau_{hL} \sim bT$.

Thus heavy component is marginal – heavy electrons more diffusively in the surrounding of light electrons. However, light electrons scatter on the heavy ones as if on a static impurity and thus light component is non-marginal. Correspondingly for the conductivities:

$$\sigma_L \sim \sigma_{Lh} \sim \frac{n_L e^2}{m_L}\tau_{Lh} \sim \frac{n_L e^2}{m_L b W_L} \sim \frac{\sigma_{min}}{b}, \qquad (10)$$

$$\sigma_h \sim \sigma_{hL} \sim \frac{\sigma_{min}}{b}\left(\frac{W_h^*}{T}\right)^2. \qquad (11)$$

with an account for Einstein relation $\frac{\partial n_h}{\partial \varepsilon} \sim \frac{W_h^*}{T}$ at high temperatures $T > W_h^*$. Hence the resistivity

$$R \sim \frac{1}{(\sigma_{Lh}+\sigma_{hL})} \sim \frac{b}{\sigma_{min}\left[1+\left(\frac{W_h^*}{T}\right)^2\right]}. \qquad (12)$$

goes on saturation in 3D case (see Fig.2). This behavior of $R(T)$ is typical for some uranium-based HF-compounds like UNi$_2$Al$_3$

### 3.2. Altshuler-Aronov effect in 2D.

In 2D case we should take into account weak-localization corrections due to quantum-mechanical backward scattering to classical Drude formulae for conductivity of the light band[21,22]: $\Delta\sigma_L/\sigma_{0L} \sim b\ln\frac{\tau_\varphi}{\tau}$, where $\sigma_{0L} = \frac{\sigma_{min}}{b}$ is classical Drude conductivity of light band, $\sigma_{min} = \frac{e^2}{\hbar}$ - Mott-Regel minimal conductivity in 2D, $\tau_\varphi = \tau_{ee} = \tilde{\tau}_{LL}$ - is decoherence time for light electrons, $\tau = \tau_{ei} = \tau_{Lh}$ and $l_{el} = v_{FL}\tau_{Lh}$ are elastic time and length, $L_\phi = \sqrt{D\tau_\varphi} = v_{FL}\sqrt{\tau_{Lh}\tilde{\tau}_{LL}}$ - is diffusive length, and $1/\tau_{Lh} \sim bW_L$ as in 3D. Correspondingly $1/\tilde{\tau}_{LL} = b^2 T$ - Altshuler-Aronov effect in "dirty" metal in 2D (electron-electron scattering time becomes marginal in dirty limit when between two subsequent scattering events for light electrons, a light electron scatters a lot of time on heavy electrons as if on almost elastic impurities, see Fig.3). Hence the conductivity of the light band: $\sigma_L = \frac{\sigma_{min}}{b}(1-b\ln\frac{W_L}{bT})$. Thus in 2D case light component has a tendency towards localization for $bT \geq W_h^*$. Moreover the additional narrowing of the heavy band and additional localization of the light band are governed for $bT \sim W_h^*$ by the same parameter $b\ln(m_h/m_L) \geq 1$.

### 3.3 Resistivity in homogeneous case in 2D.

Thus instead of desired marginal Fermi-liquid behavior at high-temperatures $T > W_h^*$ in 2D we have even more interesting behavior of resistivity $R \sim \frac{1}{(\sigma_L+\sigma_h)}$, where $\sigma_h \sim \frac{\sigma_{min}}{b}\left(\frac{W_h^*}{T}\right)^2$ as in the 3D case. Namely $R(T)$ in 2D has a maximum and then a localization tail at higher temperatures (see Fig.4). Such resistivity characteristics resembles the curve for $R(T)$ in optimally doped layred CMR-systems.

### 3.4 Superconductivity in the two – band model with one narrow band.

In the homogeneous state the leading instability in the two band model at low electron densities corresponds to p-wave pairing via enhanced Kohn-Luttinger mechanism of SC [23-27]. Namely SC critical temperature is mostly governed by the pairing of heavy electrons via polarization of light electrons. P-wave critical temperature $T_{C1}$ is strongly dependent upon relative fillings of the two bands $n_h/n_L$ and has a large and broad maximum for $n_h/n_L \sim 4$ in 2D [13,14,26,27]. For $\varepsilon_{Fh} \sim (30\text{-}50)K$ – typical for HF-compounds or semimetals (superlattices, heterostructures in 2D) $T_{C1}$ can reach *(1-5)K* which is quite nice [13,14]. The two SC gaps for heavy and light electrons are opened simultaneously below this temperature [6].

## Conclusions.

We have analyzed EPE and other mechanisms of mass-enhancement for the heavy electrons in the framework of the two-band Hubbard model with one narrow band. These mechanisms can produce the effective heavy masses $m_h^* \sim 100 m_e$ which are typical for uranium-based HF-compounds. For a large mismatch between the densities of heavy and light bands $n_h \gg n_L$ we also found a tendency towards phase-separation in 3D. We evaluate scattering times and resistivities in the homogeneous case in 3D and in 2D. Both in 3D and 2D cases at low temperatures $T < W_h^*$ the resistivity behaves in Landau Fermi-liquid fashion. At high temperatures $T > W_h^*$ the resistivity in 3D goes on saturation as in UNi$_2$Al$_3$. In 2D case due to weak-localization corrections of Altshuler-Aronov type the resistivity has a maximum and then a localization tail at higher temperatures. We analyzed the possibility of SC-transition in this model. The leading instability is towards triplet p-wave pairing and is governed by enhanced KL-mechanism of SC for pairing of heavy electrons via polarization of light electrons.

**Acknowledgements.** We are grateful to P. Fulde, Yu. Kagan, K.I. Kugel, N.V. Prokof'ev, P. Nozieres, and C.M. Varma for the numerous stimulating discussions. We acknowledge financial support of the RFBR grant № 11-02-00741.

Figure Captions

Fig. 1 The band structure in the two-band model with one narrow band. Wh and WL are the bandwidths of heavy and light electrons

Fig. 2 The resistivity characteristics R(T) in the two-band model in 3D.

Fig. 3 Multiple scattering of light particle on the heavy ones in between of the scattering of light particle on another light particle. Lφ is a diffusive length, l is elastic length

Fig. 4 Resistivity R(T) in a 2D case for the two-band model with one narrow band

# FIGURES

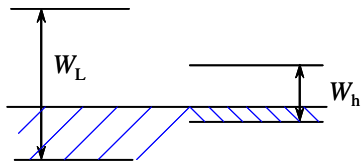

**Fig. 1**

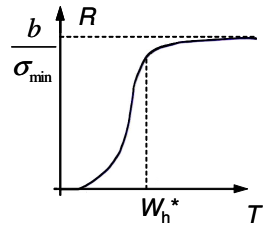

**Fig. 2**

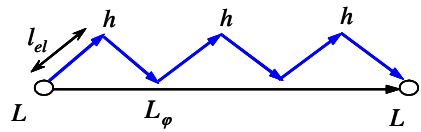

**Fig. 3**

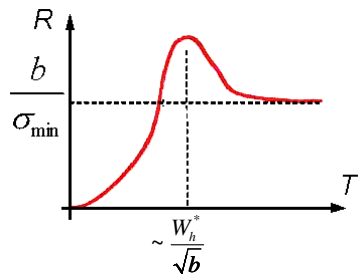

**Fig. 4**